\documentclass{article}
\pdfoutput=1

\usepackage{arxiv}

\usepackage[utf8]{inputenc} 
\usepackage[T1]{fontenc}    
\usepackage{hyperref}       
\usepackage{url}            
\usepackage{booktabs}       
\usepackage{amsfonts}       
\usepackage{nicefrac}       
\usepackage{microtype}      
\usepackage{lipsum}
\usepackage{amsmath}
\usepackage{amsthm}
\usepackage{amssymb}
\usepackage{physics}
\usepackage{graphicx}

\title{Classicalization in Derivatively Coupled Scalar Field Theories: A Feasibility Study}

\author{
  Mishkat Al Alvi \\
  Department of Physics and Astronomy\\
  Department of Mathematics and Statistics\\
  University of Minnesota Duluth\\
  Duluth, MN 55812 \\
  \texttt{alvi0016@d.umn.edu} \\
   \And
 Arshad Momen \thanks{On leave of absence from Theoretical Physics Department, University of Dhaka, Dhaka, Bangladesh.}\\
  Department of Physical Sciences\\
  Independent University, Bangladesh\\
  Bashundhara R/A, Dhaka-1212, Bangladesh \\
  \texttt{arshad@iub.edu.edu; arshad.momen@gmail.com} \\
}

\date{}

\begin{document}
\maketitle

\begin{abstract}
It has been suggested that a certain class of UV-incomplete quantum field theories can avoid unitarity violation above the cut-off energy scale by forming classical configurations at a length scale much larger than the cut-off length. This phenomenon has been named classicalization and is characterized by a length scale called classicalization radius $r_*$ which increases with energy. 
It has been argued that scalar field theories with derivative self-interactions are likely candidate for UV-completetion by classicalization and are much likely to form classicalons compared to non-classicalizing theories like $\phi^4$ scalar field theory. To look further into this claim, in this paper 2 to N particle scattering amplitude, scattering cross-section  and the amplitude of classical structure formation has been calculated and compared for a classicalizing and non-classicalizing theory. As the phenomenon of classicalization relies on creating a large number of low energy particles from high energy two particle scattering, the ratios between the scattering amplitudes and the amplitude of classical structure formation in these two cases are an indicator of the feasibility of the classicalization process. 
From our calculation, it has been observed that with the increase of energy, the ratios of the relevant quantities between classicalizing and non-classicalizing theory actually decrease which is quite contrary to the expected behaviour if classicalization is supposed to self-unitarize certain class of theories beyond cut-off energy. 
\end{abstract}


\section{Introduction}
Physics is a coordinated and logical effort to understand nature at different length scales. Physical theories most of the time has an implicit minimum distance d (maximum energy $\Lambda$) beyond which the theory looses its applicability. For example, quantum electrodynamics of electrons and photons is only applicable below the energy scale $\Lambda=2m_\mu$, twice the muon mass. Beyond this energy scale muons can pair produce themselves even if they were not present initially. So, the correct theory beyond this energy scale is the quantum electrodynamics of electron, photon and muon. One way to look at this is, below the energy scale $2m_\mu$, the degrees of freedom needed to describe the theory were electrons and photons. But when the energy scale becomes larger than $2m_\mu$, the old degrees of freedom aren't adequate anymore. We have to introduce new degrees of freedom muons, alongside the old degrees of freedom electron and photons to describe physics accurately beyond the energy scale $2m_\mu$. In effective field theory, we build up a theory using the degrees of freedom most suitable for the relevant energy scale. The degrees of freedom used are weakly interacting. However, as we increase the energy beyond some energy scale, the theory leaves its domain of applicability and must be completed by another theory which is relevant for the new energy scale. So, as we move towards the shorter-distance (higher energy) sector, an UV-completion of the theory must be performed. One of the most prominent signs for the need of UV-completion is beyond the energy scale $\Lambda$, some degrees of freedom become strongly interacting. In the standard approach dubbed as the Wilsonian approach, the UV-completion is achieved by adding in some new weakly interacting degrees of freedom. \\

Another idea that is closely connected with UV-completion is renormalization. As it has been said before, many theories come with an energy scale $\Lambda$ above which the theory looses its predictive power. A theory is said to be non-renormalizable if the dynamics of the theory at an energy far-below the cut-off is dependent on the detailed physics at energy $\Lambda$. Gravity can be sited as an example of a successful but non-renormalizable theory. However, in the case of renormalizable theories, the cut-off $\Lambda$ only influences the physics at lower energies through a small number of parameters such as mass, charge etc. of all or some of the particles involved. Once we get the values of these parameters from experiment, our theory is ready to go. One very common approach to renormalize a theory is to UV-complete the theory in a Wilsonian fashion beyond the cut-off energy scale $\Lambda$. However, not all theories can be UV-completed in this way. For some theories, there exists no Wilsonian UV-completion beyond the energy scale $\Lambda$. These theories are considered non-renormalizable and are thought to be violating unitarity above the energy scale. \\

Let us consider non-renormalizable quantum field theories with a cut-off $M_*=L_*^{-1}$. It is generally assumed that, this type of theories essentially violates unitarity above the energy scale $M_*$ and therefore is UV-incomplete. The Wilsonian approach to UV-completion recommends that, unitarity must be restored by integrating in some weakly coupled external degrees of freedom above the energy scale $M_*$. The common physical intuition that scattering processes with energies $\sqrt{s}>>M_*$ will be able to probe lengths $L<<L_*$  strengthens this opinion. Now the question logically arises, is the Wislsonian approach to UV-completion the only way to restore unitarity for UV-incomplete theories? Or the theory might possess some internal resources to prevent itself from being probed in lengths less than its cut-off length? The behaviour of gravity generally comes into mind in this aspect. Although gravity is an UV-incomplete theory, the formation of black holes prevents it from being probed in arbitrary length scales. For example, let us consider, a two-to-two graviton scattering at distance $\frac{1}{\sqrt{s}}$ and energy $\sqrt{s}$. Whenever $\sqrt{s}>>M_P$, this process will violate unitarity, because effective gravitational coupling increases with energy and finally blows up for $s>>{M_P}^2$. However, it is impossible to localize a particle pair with center-of-mass energy $\sqrt{s}>>M_P$ in a distance shorter than their corresponding Schwarzschild radius $R_s=2G_N\sqrt{s}$. This fact is insensitive to the specifics of the short-distance physics. Instead of two-to-two scattering, the scattering process is dominated by the production of black holes, which then evaporates to many low-energy particles. Instead of probing shorter distances, as we increase the energy longer length becomes important. In other words, the theory cures itself of unitarity violation by production of objects which have length scales of the classical type. This process may be referred to as classicalization. Although, we have used here an example from gravity, it might be possible for UV-incomplete quantum field theories to pursue this path for restoring unitarity beyond the cut-off scale. In this paper, we will have a closer look at the classicalization process and will try to understand whether it is a viable route for UV-completion of UV-incomplete theories. \\

Classicalization can be defined as the phenomenon where an UV-incomplete theory UV completes itself by creating a classical configuration (which we might call classicalon) sourced by Noether type charges (like energy, momentum etc.)- whenever the theory has to deal with scattering occurring at CM energies above the cut-off energy scale \cite{dvali2011uv}. The size of the classical configuration is dictated by the classicalization radius $r_*$ , which is a model-dependent function of the energy of the source $\sqrt{s}$. The key feature of this classicalization length scale $r_*$ is: for trans cut-off energies $\sqrt{s}>>M_*$ the classicalization radius always exceeds the cut-off length-scale of the theory $L_*$. As a result, no matter how energetic some scattering event is, the theory prevents itself from getting probed beyond the cut-off length scale $L_*$. Although this phenomenon has a strong resemblance with black hole formation in gravity theories, other fancy aspects of gravity like event horizon, information problem etc. are not relevant in this case. One question naturally arises though. In general classical object formation like topological solitons is exponentially suppressed in two-particle scattering, so how classicalization can even happen? The answer is classicalons are sourced by Noether type charges instead of topological charges, so whenever localization of energy-momentum happens under the classicalization radius $r_*$, classicalons are bound to be produced, according to the original suggestion \cite{dvali2011uv}. Now, the question that needs to be answered is: Does classicalization really happen? In order to answer this question, a comparative study of the theories which are probable candidate for displaying classicalizing behavior and the ones that are UV-complete and do not display classicalization was done in \cite{dvali2011dynamics}. To accomplish this, we need to develop a universal language first, to judge both type of theories on the same pedestal. In our paper, the scattering cross-section has been defined as a geometric cross-section characterized by the radius $r_*$, conforming to the convention used in \cite{dvali2011dynamics}. $r_*$ radius can be understood as the shortest distance down to which scattering waves can propagate freely without experiencing any significant interaction and distortion. The scattering cross-section can be written as 

\begin{equation}
\sigma(r_*)\sim r_*(\sqrt{s})^2
\end{equation}  

This is applicable for both classicalizing and non-classicalizing theories. The difference lies in the dependence of $r_*$ on energy. In non-classicalizing theories $r_*$ decreases with $\sqrt{s}$ allowing scattering experiments to probe lesser and lesser distances with increasing energy. In classicalizing theories, the opposite happens, $r_*$ is actually proportional to $\sqrt{s}$, making it impossible to probe distances shorter than $L_*$. As it has been shown in \cite{dvali2011dynamics}, derivatively coupled theories such as $\frac{1}{M_*^4}((\partial_\mu\phi)^2)^2$ are possible candidates for classicalization because they are sourced by energy and their scattering amplitude might follow our criteria of increasing with energy. As it has been shown in \cite{dvali2011dynamics}, for derivatively self-interacting scalar field, the $r_*$ radius is given by

\begin{equation}
r_*(\sqrt{s})=L_*(L_*\sqrt{s})^{\frac{1}{3}}
\end{equation}

So, with the increase of energy, the $r_*$ radius actually increases. This theory might show classicalizing behaviour, at least the initial indication and the primary analysis says so. In this paper, we will try to analyze further in this direction i.e. whether this derivatively coupled theory has a greater edge in classicalon formation than $\phi^4$ theory. As it has been shown that, $\phi^4$ theories can not produce classicalon \cite{dvali2011dynamics}, a relative comparison between the two theories might shed some light on the classicalizability of the derivatively coupled one. For classicalization it is necessary to go through a $2\rightarrow N$ scattering, if two particle scattering with a CM energy $E>>M_*$ is going to form a classical configuration consisting of N particles. The value of N has been estimated as $N=(\frac{\sqrt{s}}{M_*})^{\frac{4}{3}}$ in \cite{dvali2017strong}. So, $2\rightarrow N$ scattering amplitude and scattering cross-section for $\frac{1}{M_*^4}((\partial_\mu\phi)^2)^2$ theory should increase with energy at a faster rate than $\phi^4$ theory, as the greater scattering energy is, the more probable classicalon formation should become. The method we have used is the direct evaluation of a subset of Feynmann diagrams to get a lower bound on the scattering amplitude. From this amplitude, scattering cross-section were calculated using the relativistic phase space approximation. Finally, the amplitude of formation of classical structure was calculated for both the theories. It has been showed that, with the increase of energy the ratio of the scattering amplitude , scattering cross-section and amplitude of classical structure formation calculated for $\frac{1}{M_*^4}((\partial_\mu\phi)^2)^2$ and $\phi^4$ theory actually decreases, quite contrary to our original expectation. It can be seen from this result, as we increase the scattering energy $2\rightarrow N$ scattering and classical structure formation becomes more favourable for $\phi^4$ theory than its derivatively coupled counterpart. As $\phi^4$ theory isn't expected to produce classicalons, it means that it is even more unlikely for $\frac{1}{M_*^4}((\partial_\mu\phi)^2)^2$  theory to produce them. From our analysis, it is apparent that  $\frac{1}{M_*^4}((\partial_\mu\phi)^2)^2$ theory actually performs worse in classicalon formation than $\phi^4$ theory in the specific case we have considered.

\section{A Tale of Two Theories: Scattering and Classical Configuration Formation}

Dvali et al. tried to show in their analysis that, only certain class of theories like the derivatively coupled scalar $\frac{1}{M_*^4}((\partial_\mu \phi)^2)^2$ theory shows classicalizing behavior, whereas the likes of $\phi^4$ theories don't \cite{dvali2011dynamics}. Whenever a scattering happens with a center of mass energy greater than the cut-off energy scale, the classicalizing theories are supposed to form a classical configuration with a high occupation number. However, no specific dynamic mechanism has been suggested. In this section, we would like to calculate the scattering amplitude and scattering cross-section of the $2\rightarrow N$ scattering for two theories, $\phi^4$ and $\frac{1}{M_*^4}((\partial_\mu \phi)^2)^2$. The scattering amplitude and scattering cross-section should be higher for the theories which are supposed to go through classicalization, as to protect unitarity they must form a lot of soft quanta to distribute the high CM-energy involved in the scattering. The amplitude for forming a classical configuration will also be calculated for both theories, so that we can get a clearer estimation about whether $\frac{1}{M_*^4}((\partial_\mu \phi)^2)^2$ will go through classicalization or not. The relevant quantities for $\phi^4$ theory were first calculated in \cite{goldberg1990breakdown}, \cite{cornwall1993functional}. The method of \cite{goldberg1990breakdown} will be used for $\frac{1}{M_*^4}((\partial_\mu \phi)^2)^2$ theory.

\subsection{$\phi^4$ Theory}  

In this section, the scattering $q\bar{q}\rightarrow nH$ will be considered, where $H$ is the scalar Higgs field. The Lagrangian for the theory can be written as,

\begin{equation}
\mathcal{L}=\frac{1}{2}(\partial_\mu H)^2-\frac{\lambda_{k+1}}{(k+1)!}H^{k+1}
\end{equation}

For all the discussions and equations afterward in this section, $k=3$ will be used. We will be considering the case where the energies of the final state particles are equal. Let us define, $n_{max}=\frac{E}{m_H}$ and $E=\sqrt{\hat{s}}$. We are selecting this scenario because this will be a reasonable estimation for an average matrix element over all of phase space when the number of particles n is greater than the limit $n>\frac{1}{2}n_{max}$. The most important source of contribution is the symmetrically branching tree as shown in the figure[1]. Although the other trees will contribute, as they will add coherently, the symmetrically branching tree remains the most interesting and dominant case for understanding the essential behaviour. So, we will get a lower bound on the amplitude by calculating the symmetrically branching tree of figure[1]. After the counting of diagrams, we will set all spatial momenta to zero to keep the lower bound, which is allowed by the timelike nature of all the energy denominators. This will be a source of added simplification. \\

\begin{figure}
\begin{center}
\includegraphics[scale=0.5]{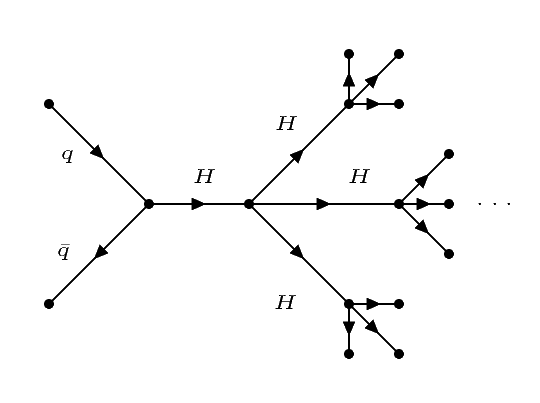}
\caption{The Symmetric Tree Net}
\end{center}
\end{figure}

The amplitude corresponding to the figure can be written as
\begin{equation}
\mathcal{A}_n(E)=g_Y\bar{u}v\frac{1}{E^2}\bar{\mathcal{M}_n}C_n
\end{equation}

Here $\bar{\mathcal{M}_n}$ is the amplitude for a single permutation of final state particles and $C_n$ is the number of topologically nonequivalent diagrams. From the figure, we can see that $n=k^p$. Now, we can write the value of $\bar{\mathcal{M}_n}$ as

\begin{equation}
\begin{split}
\bar{\mathcal{M}_n} & = \lambda_{k+1} \frac{\lambda_{k+1}^k\lambda_{k+1}^{k^2}\dots \lambda_{k+1}^{k^{p-1}}}{[(\frac{E}{k})^k(\frac{E}{k^2})^{k^2}\dots (\frac{E}{k^{p-1}})^{k^{p-1}}]^2} \\
& = \frac{\lambda_{k+1}^{1+k+k^2+\dots +k^{p-1}}}{\frac{(E^2)^{k+k^2+\dots +k^{p-1}}}{(k^2)^{k+2k^2+\dots (p-1)k^{p-1}}}} \\
& \approx \lambda_{k+1}(\frac{\lambda_{k+1}}{E^2})^{\frac{n-k}{k-1}}k^{\frac{2kn[p(1-\frac{1}{k})-1]}{(k-1)^2}} \\
& = \lambda_{k+1}(\frac{E^2}{\lambda_{k+1}})^{\frac{k}{k-1}}(\frac{\sqrt{\lambda_{k+1}}}{k^{\frac{k}{k-1}}}\frac{n}{E})^{\frac{2n}{k-1}}
\end{split}
\end{equation}

Now we will have to calculate the combinatoric factor. This can be done by partitioning the n particles (labeled by their momenta before setting them to zero) into distinct groups of $\frac{n}{k}$ particles, each of these is then divided into distinct groups of $\frac{n}{k^2}$ particles and so on. So the combinatoric factor can be written as

\begin{equation}
\begin{split}
C_n & = \frac{n!}{k![(\frac{n}{k})!]^k}(\frac{(\frac{n}{k})!}{k![(\frac{n}{k^2})!]^k})^k(\frac{(\frac{n}{k^2})!}{k![(\frac{n}{k^3})!]^k})^{k^2} \dots (\frac{k!}{k!(1!)^k})^{k^{p-1}} \\
& = \frac{n!}{(k!)^{\frac{n-1}{k-1}}}
\end{split}
\end{equation}

So, if we ignore numerical factors of $O(1)$, the amplitude for the symmetric tree can be written as

\begin{equation}
\begin{split}
\mathcal{M}_n & = \bar{\mathcal{M}_n}C_n \\
& = \lambda_{k+1}(\frac{E^2}{\lambda_{k+1}})^{\frac{k}{k-1}}(\frac{\sqrt{\lambda_{k+1}}}{k^{\frac{k}{k-1}}}\frac{n}{E})^{\frac{2n}{k-1}}\frac{n!}{(k!)^{\frac{n-1}{k-1}}}
\end{split}
\end{equation}

It must be kept in mind that, this is a lower bound on the amplitude. A similar result has also been derived by Cornwall\cite{cornwall1993functional}, albeit by different method. If the number of particles is such that $\frac{E}{nM_H}$ is appropriate for relativistic phase space,  then we can calculate the lower bound on scattering cross-section.

\begin{equation}
\begin{split}
\sigma_{q\bar{q}\rightarrow nH} & \approx \frac{g_Y^2}{\hat{s}^2} |\mathcal{M}_n|^2 \rho_n(E) \\
& = \frac{g_Y^2}{\hat{s}^2}\lambda_{k+1}^2(\frac{E^2}{\lambda_{k+1}})^{\frac{2k}{k-1}}(\frac{\lambda_{k+1}}{k^{\frac{2k}{k-1}}}\frac{n^2}{E^2})^{\frac{2n}{k-1}}\frac{(n!)^2}{(k!)^{\frac{2(n-1)}{(k-1)}}}\frac{1}{(n!)^3}\frac{\hat{s}^{n-2}}{(4\pi)^{2n-3}} \\
& = \frac{g_Y^2}{\hat{s}^2}\lambda_{k+1}^2(\frac{\hat{s}}{\lambda_{k+1}})^{\frac{2k}{k-1}}(\frac{\lambda_{k+1}}{k^{\frac{2k}{k-1}}}\frac{n^2}{\hat{s}})^{\frac{2n}{k-1}}\frac{(n!)^2}{(k!)^{\frac{2(n-1)}{k-1}}}\frac{1}{(n!)^3}\frac{\hat{s}^{n-2}}{(4\pi)^{2n-3}} \\
& = g_Y^2\frac{\lambda_{k+1}^{n-1}n^{2n}}{3^{3n}(n!)6^{n-1}(4\pi)^{2n-3}}\frac{1}{\hat{s}}
\end{split}
\end{equation}

Here, $\rho_n\approx\frac{1}{(n!)^3}\frac{\hat{s}^{n-2}}{(4\pi)^{2n-3}}$ and $\hat{s}=E^2$ is the square of the CM energy. \\

Now, the amplitude for the formation of classical configuration will be calculated. Let us consider that, $H_c$ is a classical Higgs field with a large number of particles n. This state can be described in momentum space as a Fock state $\ket{H_c}$ with

\begin{equation}
\bra{k_1k_2\dots k_n}\ket{H_c}= e^{-\frac{\bar{n}}{2}}\prod_{i=1}^{n}f(k_i)
\end{equation}

Here,
\begin{equation}
\sum_{n}\frac{1}{n!}\int\prod_{i=1}^{n}\frac{d^3\bold{k_i}}{(2\pi)^3}\ket{k_1\dots k_n}\bra{k_1\dots k_n}=1
\end{equation}

and

\begin{equation}
\int\frac{d^3\bold{k}}{(2\pi)^3}=\bar{n} 
\end{equation}

So, the lower bound of the amplitude of generating the classical field $H_c$ from the incoming $q\bar{q}$ scattering can be written as

\begin{equation}
\begin{split}
\bra{H_c}T\ket{q\bar{q}} & \approx e^{-\frac{\bar{n}}{2}}\sum_{n}\frac{1}{n!}\int\prod_{i=1}^n\frac{d^3\bold{k_i}}{(2\pi)^3}f(k_i)\mathcal{M}_n(k_1\dots k_n)\\
& \approx e^{-\frac{\bar{n}}{2}}\sum_{n}(H_c(0))^n\frac{1}{n!}(\frac{\sqrt{\lambda}}{a}\frac{n}{\sqrt{\hat{s}}})^n n!
\end{split}
\end{equation}

Here, $a=3\sqrt{3}\approx5.2$ in Goldberg's calculation\cite{goldberg1990breakdown} and $a=2.92$ in Cornwal's estimate\cite{cornwall1993functional}. So, if we consider $\sqrt{\hat{s}}\approx \hat{n}m_H$, the overlap becomes large for,

\begin{equation}
\begin{split}
H_c(0)\geq \frac{a\sqrt{\hat{s}}}{\sqrt{\lambda}n} \\
\geq \frac{am_H}{\sqrt{\lambda}}
\end{split}
\end{equation}

So, from this result, it is clear that, certain classical configuration can be produced in unsuppressed manner in high energy two-particle collisions, even in the case of $\phi^4$ theory. If the classicalization conjecture is true, the performance of derivatively coupled scalar theories should be even better in producing classical configuration from two particle scattering. We will consider this in the next case.

\subsection{$\frac{1}{M_*^4}((\partial_\mu\phi)^2)^2$ Theory}

We will consider the same $q\bar{q}\rightarrow nH$ scattering in this case, too. The Lagrangian for the derivatively coupled theory can be written as \cite{dvali2017strong}
\begin{equation}
\mathcal{L}=\frac{1}{2}(\partial_\mu H)^2+\frac{1}{4M_*^4}((\partial_\mu H)^2)^2
\end{equation}

The tree diagram in this case will be the one showed in figure[1]. We will use the same method that has been used in the previous section for $\phi^4$ theory. The energies of the final states will be equal and the other considerations will also be applicable in this case. The most important contribution will come from the symmetrical branching tree of figure[1] and by evaluating that amplitude, we will get a lower bound on the scattering amplitude. We will define $n_{max}=\frac{E}{m_H}$ and $E=\sqrt{\hat{s}}$ and when n is not much smaller than $n_{max}$, this will be a reasonable approximation for a generic matrix element over all of phase space. \\

The amplitude can be written as
\begin{equation}
\mathcal{A}_n(E)=g_Y\bar{u}v\frac{1}{E^2}\bar{\mathcal{M}_n}C_n
\end{equation}

Just like $\phi^4$ theory, $\bar{\mathcal{M}_n}$ is the amplitude for a single permutation of final state particles and $C_n$ is the number of topologically nonequivalent diagrams. From the figure, we can see that $n=k^p$. For the equations below, we will use $\lambda_k=\frac{1}{M_*^4}$. Now,

\begin{equation}
\begin{split}
\mathcal{\bar{M}}_n & =\lambda_kE^4[\frac{\lambda_k(\frac{E}{k})^4}{(\frac{E}{k})^2}]^k[\frac{\lambda_k(\frac{E}{k^2})^4}{(\frac{E}{k^2})^2}]^{k^2}\dots [\frac{\lambda_k(\frac{E}{k^{p-1}})^4}{(\frac{E}{k^{p-1}})^2}]^{k^{p-1}} \\
 & = E^4\lambda_k[\lambda_k(\frac{E}{k})^2]^k[\lambda_k(\frac{E}{k^2})^2]^{k^2}\dots [\lambda_k(\frac{E}{k^{p-1}})^2]^{k^{p-1}} \\
 & = E^4(\lambda_k\lambda_k^k\lambda_k^{k^2}\dots \lambda_k^{k^{p-1}})\frac{(E^2)^k(E^2)^{k^2}\dots (E^2)^{k^{p-1}}}{(k^2)^k(k^2)^{2k^2}\dots (k^2)^{(p-1)k^{p-1}}} \\
 & =E^4(\lambda_k)^{1+k+k^2+\dots +k^{p-1}}\frac{(E^2)^{k+k^2+\dots +k^{p-1}}}{(k^2)^{k+2k^2+\dots (p-1)k^{p-1}}}\\
 & =E^4(\lambda_k)^{\frac{n-1}{k-1}}\frac{(E^2)^{\frac{n-k}{k-1}}}{(k^2)^{\frac{(pk-p-k)n+k}{(k-1)^2}}} \\
\end{split}
\end{equation}

The combinatoric factor remains the same as $\phi^4$ theory, because the structure of the symmetric branching tree is same for both cases. So,

\begin{equation}
C_n=\frac{n!}{(k!)^{\frac{n-1}{k-1}}}
\end{equation}

The value of $M_n$ is
\begin{equation}
\begin{split}
\mathcal{M}_n & =\bar{\mathcal{M}_n}C_n \\
& = \frac{n!}{(k!)^{\frac{n-1}{k-1}}}E^4(\lambda_k)^{\frac{n-1}{k-1}}\frac{(E^2)^{\frac{n-k}{k-1}}}{(k^2)^{\frac{(pk-p-k)n+k}{(k-1)^2}}}
\end{split}
\end{equation}

This is the lower bound for the scattering amplitude, just like the $\phi^4$ case. Now, 
\begin{equation}
\begin{split}
|\mathcal{M}_n|^2 & = \frac{(n!)^2}{(k!)^{\frac{2(n-1)}{k-1}}}(E^4)^2(\lambda_k)^{\frac{2(n-1)}{k-1}}\frac{(E^2)^{\frac{2(n-k)}{k-1}}}{(k^2)^{\frac{2(pk-p-k)n+2k}{(k-1)^2}}} \\
& \approx \frac{(n!)^2}{(k!)^{\frac{2(n-1)}{k-1}}}(E^4)^2(\lambda_k)^{\frac{2(n-1)}{k-1}}\frac{(E^2)^{\frac{2(n-k)}{k-1}}}{(k^2)^{\frac{2(k-1)pn+2k}{(k-1)^2}}} \\
& =\frac{(n!)^2}{6^{n-1}}(E^4)^2(\lambda_k)^{n-1}\frac{(E^2)^{n-3}}{(k^2)^{pn}3^3} \\
& =\frac{(n!)^2}{6^{n-1}}(\lambda_k)^{n-1}\frac{(E^2)^{n+1}}{n^{2n}3^3}
\end{split}
\end{equation}

If the number of particles n is such that $\frac{E}{nM_H}$ is suitable for using relativistic phase space, then the lower bound on cross-section will be given by
\begin{equation}
\begin{split}
\sigma_{q\bar{q}\rightarrow nH} & \approx \frac{g_Y^2}{\hat{s}^2}|\mathcal{M}_n|^2\rho_n(E) \\
& = \frac{g_Y^2}{\hat{s}^2}\frac{(n!)^2}{6^{n-1}n^{2n}3^3}(\lambda_k)^{n-1}(\hat{s})^{n+1}\frac{1}{(n!)^3}\frac{\hat{s}^{n-2}}{(4\pi)^{2n-3}}\\
& = g_Y^2\frac{1}{(n!)6^{n-1}n^{2n}3^3}(\lambda_k)^{n-1}\frac{(\hat{s})^{2n-3}}{(4\pi)^{2n-3}}
\end{split}
\end{equation}

Just as in the case of $\phi^4$ theory, we will now calculate the amplitude of generating a classical field configuration $H_c$ from high energy two-particle scattering. Let us consider, $H_c$ is a classical Higgs field which can be describes in momentum space as a Fock state $\ket{H_c}$ with

\begin{equation}
\bra{k_1k_2\dots k_n}\ket{H_c}=e^{-\frac{\bar{n}}{2}}\prod_{i=1}^n f(k_i)
\end{equation}

Here,
\begin{equation}
\sum_{n}\frac{1}{n!}\int\prod_{i=1}^{n}\frac{d^3\bold{k_i}}{(2\pi)^3}\ket{k_1\dots k_n}\bra{k_1\dots k_n}=1
\end{equation}

and

\begin{equation}
\int\frac{d^3\bold{k}}{(2\pi)^3}=\bar{n} 
\end{equation}

So, the lower bound on the amplitude for generating classical field $H_c$ from the incoming state $q\bar{q}$ can be estimated as

\begin{equation}
\begin{split}
\bra{H_c}T\ket{q\bar{q}} & \approx e^{-\frac{\bar{n}}{2}}\sum_{n}\frac{1}{n!}\int\prod_{i=1}^n\frac{d^3\bold{k_i}}{(2\pi)^3}f(k_i)\mathcal{M}_n(k_1\dots k_n)\\
\end{split}
\end{equation}

Now, 
\begin{equation}
\begin{split}
\mathcal{M}_n & =\bar{\mathcal{M}_n}C_n 
= \frac{n!}{(k!)^{\frac{n-1}{k-1}}}E^4(\lambda_k)^{\frac{n-1}{k-1}}\frac{(E^2)^{\frac{n-k}{k-1}}}{(k^2)^{\frac{(pk-p-k)n+k}{(k-1)^2}}} \\
& \approx \frac{(n!)}{6^{\frac{n-1}{2}}}(\lambda_k)^{\frac{n-1}{2}}\frac{E^4E^{n-3}}{(k^2)^{\frac{(k-1)pn}{(k-1)^2}}(k^2)^{\frac{k}{(k-1)^2}}} \\
& \approx \frac{n!}{(\sqrt{6})^n}(\sqrt{\lambda_k})^{n-1}\frac{(E)^{n+1}}{(k^2)^{\frac{pn}{2}}(k^2)^{\frac{3}{4}}}\\
& = \frac{n!}{(\sqrt{6})^n}\frac{(\sqrt{\lambda_k})^n}{\sqrt{\lambda_k}}\frac{E^nE}{n^n3^{\frac{3}{2}}} \\
& \approx (n!)(\frac{E}{\sqrt{\lambda_k}})(\frac{E\sqrt{\lambda_k}}{n\sqrt{6}})^n \\
& = n!(EM_*^2)(\frac{1}{\sqrt{6}n}\frac{E}{M_*^2})^n \\
& \approx n! (\frac{1}{\sqrt{6}n}\frac{\sqrt{s}}{M_*^2})^n
\end{split}
\end{equation}

So, if we use this value of $\mathcal{M}_n$ then we get,
\begin{equation}
\bra{H_c}T\ket{q\bar{q}}\approx e^{-\frac{n}{2}}\sum_{n}(H_c(0))^n\frac{1}{n!}(\frac{1}{\sqrt{6}n}\frac{E}{M_*^2})^nn!
\end{equation}
This is the lower bound on the amplitude of classical structure formation for $\frac{1}{M_*^4}((\partial_\mu\phi)^2)^2$ theory. \\
Now, this overlap will become large if 
\begin{equation}
|H_c(0)|\geq \frac{\sqrt{6}nM_*^2}{E}
\end{equation}
So, certain classical structure can be generated in unsuppressed manner in high-energy two-particle collision.

\subsection{To Classicalize or Not To Classicalize}
We have calculated the lower bound on scattering amplitude, scattering cross-section and amplitude of classical structure formation for both of the theories. As it has been speculated before \cite{dvali2011dynamics} that, $\phi^4$ theory does not show classicalizing behaviour, but $\frac{1}{M_*^4}((\partial_\mu\phi)^2)^2$ theory does; the ratio of the magnitudes of the corresponding quantities (i.e. scattering amplitude, scattering cross-section and amplitude of classical structure formation ) of both the theories will shed some more light on the classicalization process.  

Now if we take the ratio of the scattering amplitude of $\frac{1}{M_*^4}((\partial_\mu\phi)^2)^2$ and $\phi^4$ theory, then we get
\begin{equation}
\begin{split}
& E^4(\lambda_k)^{\frac{n-1}{k-1}}\frac{(E^2)^{\frac{n-k}{k-1}}}{(k^2)^{\frac{(pk-p-k)n+k}{(k-1)^2}}}\frac{1}{\lambda_{k+1}}(\frac{\lambda_{k+1}}{E^2})^{\frac{k}{k-1}}(\frac{k^{\frac{k}{k-1}}}{\sqrt{\lambda_{k+1}}}\frac{E}{n})^{\frac{2n}{k-1}} \\
& \approx E^4(\lambda_k)^{\frac{n-1}{2}}\frac{(E^2)^{\frac{n-3}{2}}}{(k^2)^{\frac{p(k-1)n}{(k-1)^2}}(k^2)^{\frac{k}{(k-1)^2}}}\frac{1}{\lambda_{k+1}}(\frac{\lambda_{k+1}}{E^2})^{\frac{3}{2}}(\frac{k^{\frac{3}{2}}}{\sqrt{\lambda_{k+1}}}\frac{E}{n})^n \\
& = E^4 (\frac{1}{M_*^4})^{\frac{n-1}{2}}\frac{E^{n-3}}{n^n 3^{\frac{3}{2}}}\frac{\lambda_{k+1}^{\frac{1}{2}}}{E^3}\frac{3^{\frac{3n}{2}}}{\lambda_{k+1}^{\frac{n}{2}}}\frac{E^n}{n^n} \\
& = (\frac{1}{M_*^2})^{n-1}\frac{(E)^{n+1}}{n^n3^{\frac{3}{2}}}\frac{1}{(\sqrt{\lambda_{k+1}})^{n-1}}\frac{3^{\frac{3n}{2}}}{n^n}E^{n-3} \\
& = (\frac{1}{\sqrt{\lambda_{k+1}}M_*^2})^{n-1}\frac{E^{2n-2}}{n^{2n}}(3^{\frac{3}{2}})^{n-1} \\
& = (\frac{1}{\sqrt{\lambda_{k+1}}M_*^2})^{n-1}\frac{(E^2)^{n-1}}{(n^2)^{n-1}n^2}(3^{\frac{3}{2}})^{n-1}\\
& = \frac{1}{n^2}(\frac{3^{\frac{3}{2}}}{\sqrt{\lambda_{k+1}}}\frac{1}{n^2}\frac{E^2}{M_*^2})^{n-1}
\end{split}
\end{equation}

Now, we need an estimation for the number of particles $n$ in the classicalization process. If we follow Gia Dvali's calculation \cite{dvali2017strong}, $n$ can be written as
\begin{equation}
n=(\frac{E}{M_*})^{\frac{4}{3}}
\end{equation}

If we use this value of $n$ in equation(102), then the ratio of the scattering amplitudes can be written as
\begin{equation}
\begin{split}
& \frac{1}{n^2}(\frac{3^{\frac{3}{2}}}{\sqrt{\lambda_{k+1}}}\frac{M_*^{\frac{8}{3}}}{E^{\frac{8}{3}}}\frac{E^2}{M_*^2})^{n-1} \\
& = \frac{1}{n^2}(\frac{3^{\frac{3}{2}}}{\sqrt{\lambda_{k+1}}}\frac{M_*^{\frac{2}{3}}}{E^{\frac{2}{3}}})^{n-1}
\end{split}
\end{equation}

One of the most important conditions for classicalization to happen is $E>>M_*$. So, the more we increase the CM energy, the smaller the amplitude of $\frac{1}{M_*^4}((\partial_\mu\phi)^2)^2$ becomes compared to that of $\phi^4$ theory. Contrary to Dvali et al's claims, the $\frac{1}{M_*^4}((\partial_\mu\phi)^2)^2$ theory is actually less prone to go through a $2\rightarrow n$ scattering than $\phi^4$ theory, which is essential for classicalization. \\

Goldberg\cite{goldberg1990breakdown} estimated the value of $n$ from a different perspective. According to \cite{goldberg1990breakdown}, the value of $n$ should be close to $n_{max}=\frac{E}{m_H}$, but far enough below it to justify the use of relativistic phase space. If we use $n=\frac{2}{3}\frac{E}{m_H}$ like Goldberg, then the ratio of scattering amplitude can be written as

\begin{equation}
\begin{split}
& \frac{1}{n^2}(\frac{3^{\frac{3}{2}}}{\sqrt{\lambda_{k+1}}}\frac{1}{\frac{4}{9}\frac{E^2}{m_H^2}}\frac{E^2}{M_*^2})^{n-1} \\
& = \frac{1}{n^2}(\frac{3^{\frac{3}{2}}}{\sqrt{\lambda_{k+1}}}\frac{9m_H^2}{4E^2}\frac{E^2}{M_*^2})^{n-1} \\
& = \frac{1}{n^2}(\frac{3^{\frac{3}{2}}}{\sqrt{\lambda_{k+1}}}\frac{9m_H^2}{4M_*^2})^{n-1} \\
\end{split}
\end{equation}

As the cut-off energy $M_*$ is obviously higher than the Higgs mass $m_H$, this ratio is less than one. And as we can see from the $n$ exponent, the value actually decreases with the increase of $n$. So, even in this estimate the $\phi^4$ theory amplitude is higher than the $\frac{1}{M_*^4}((\partial_\mu\phi)^2)^2$ theory amplitude and with increasing value of $n$, this difference only increases. This is the exact opposite of the behavior that would be shown if $\frac{1}{M_*^4}((\partial_\mu\phi)^2)^2$ theory were better suited for classicalization than $\phi^4$. \\

Now, let's check the ratio between scattering cross-section for $2\rightarrow n$ scattering between $\frac{1}{M_*^4}((\partial_\mu\phi)^2)^2$ and $\phi^4$ theory. If we take the ratio of scattering cross-section for both theories, 
\begin{equation}
\frac{1}{n^4}(\frac{3^3}{\lambda_{k+1}}\frac{1}{n^4}\frac{E^4}{M_*^4})^{n-1}
\end{equation}

If we use the value of $n$ suggested in \cite{dvali2017strong}, then 
\begin{equation}
\begin{split}
& \frac{1}{n^4}(\frac{3^3}{\lambda_{k+1}}\frac{1}{(\frac{E}{M_*})^{\frac{16}{3}}}\frac{E^4}{M_*^4})^{n-1}\\
& = \frac{1}{n^4}(\frac{3^3}{\lambda_{k+1}}\frac{M_*^{\frac{16}{3}}}{E^{\frac{16}{3}}}\frac{E^4}{M_*^4})^{n-1}\\
& = \frac{1}{n^4}(\frac{3^3}{\lambda_{k+1}}\frac{M_*^{\frac{4}{3}}}{E^{\frac{4}{3}}})^{n-1}
\end{split}
\end{equation}

Just as in the case of ratio of scattering amplitude, the scattering cross-section ratio also shows the same pattern. If we use value of $n$ specified above, we see that, the scattering cross-section is actually less for $\frac{1}{M_*^4}((\partial_\mu\phi)^2)^2$ than $\phi^4$ theory, due to the fact that $E>>M_*$. Moreover, as the cm energy of the scattering increases, the ratio decreases, contrary to the analysis of \cite{dvali2011dynamics} that  $\frac{1}{M_*^4}((\partial_\mu\phi)^2)^2$ is a better choice for classicalization than $\phi^4$ theory. \\

Now, we will use another estimation for $n$ given by Goldberg\cite{goldberg1990breakdown}, $n=\frac{2}{3}\frac{E}{m_H}$. If we put this value in equation($106$), the ratio will look like,
\begin{equation}
\begin{split}
& \frac{1}{n^4}(\frac{3^3}{\lambda_{k+1}}\frac{1}{(\frac{2}{3})^4\frac{E^4}{m_H^4}}\frac{E^4}{M_*^4})^{n-1} \\
& = \frac{1}{n^4}(\frac{3^3}{\lambda_{k+1}}\frac{81m_H^4}{16M_*^4})^{n-1}
\end{split}
\end{equation}

As the value of the cut-off energy scale $M_*$ is higher than the Higgs mass $m_H$, this quantity is also less than 1. And the more we increase the CM energy $E$, the greater will be the value of $n$, so the ratio actually becomes lesser in magnitude with increasing energy. This behavior confirms the previous trends calculated above and shows that $\frac{1}{M_*^4}((\partial_\mu\phi)^2)^2$ does indeed have a worse performance in scattering cross-section compared to $\phi^4$ theory. \\

The last parameter we are going to check is the amplitude of classical structure formation. Here, we are gonna check the ratio of the n-th term from the summation of equation ($26$) and ($12$), to check whether they also follow the trend seen previously. 

\begin{equation}
\begin{split}
& (\frac{1}{\sqrt{6}}\frac{E}{nM_*^2})^n(\frac{a}{\sqrt{\lambda_{k+1}}}\frac{E}{n})^n \\
& = (\frac{1}{\sqrt{6}}\frac{a}{\lambda_{k+1}}\frac{E^2}{n^2M_*^2})^n \\
& \approx(\frac{1}{\sqrt{\lambda_{k+1}}}\frac{1}{n^2}\frac{E^2}{M_*^2})^n \\ 
\end{split}
\end{equation}

Just as before, we will use two values of $n$ to see how this term scales with energy. If we use the Dvali's  expression for $n$, then

\begin{equation}
\begin{split}
& (\frac{1}{\sqrt{\lambda_{k+1}}}\frac{1}{(\frac{E}{M_*})^{\frac{8}{3}}}\frac{E^2}{M_*^2})^n \\
& = (\frac{1}{\sqrt{\lambda_{k+1}}}\frac{M_*^{\frac{8}{3}}}{E^{\frac{8}{3}}}\frac{E^2}{M_*^2})^n \\
& = (\frac{1}{\sqrt{\lambda_{k+1}}}\frac{M_*^{\frac{2}{3}}}{E^{\frac{2}{3}}})^n \\
\end{split}
\end{equation}

As $E>>M_*$ for classicalization, the more we increase the energy, the less this ratio becomes. So, the amplitude for classical structure formation will be lesser for $\frac{1}{M_*^4}((\partial_\mu\phi)^2)^2$ theory than $\phi^4$ theory. If $\frac{1}{M_*^4}((\partial_\mu\phi)^2)^2$ theory were more prone to classicalization than $\phi^4$ theory, then the behaviour would have been the exact opposite. \\

Now, we will evaluate the ratio with $n=\frac{2}{3}\frac{E}{m_H}$.

\begin{equation}
\begin{split}
& (\frac{1}{\sqrt{\lambda_{k+1}}}\frac{1}{\frac{4}{9}\frac{E^2}{m_H^2}}\frac{E^2}{M_*^2})^n \\
& = (\frac{1}{\sqrt{\lambda_{k+1}}}\frac{9m_H^2}{4E^2}\frac{E^2}{M_*^2})^n \\
& = (\frac{1}{\sqrt{\lambda_{k+1}}}\frac{9}{4}\frac{m_H^2}{M_*^2})^n
\end{split}
\end{equation}

As $M_*>m_H$, even in this case the amplitude for classical configuration formation is less for $\frac{1}{M_*^4}((\partial_\mu\phi)^2)^2$ theory than $\phi^4$ theory and this ratio decreases continuously with increasing energy. This result fits well with the previous results we have obtained. \\

From this analysis, we can safely say that contrary to some other works in classicalization like \cite{dvali2011dynamics}, $\frac{1}{M_*^4}((\partial_\mu\phi)^2)^2$ theory is less suitable for classicalization than $\phi^4$ theory. Although this analysis doesn't show whether or not classicalization happens, this definitely shows that derivatively coupled theories might not have a special privilege in classical structure formation. As we can clearly see, with the increase of energy, all the relevant quantities for $2\rightarrow N$ scattering and classicalization like scattering amplitude, scattering cross-section and amplitude of classical structure formation clearly decrease for $\frac{1}{M_*^4}((\partial_\mu\phi)^2)^2$ theory at a rate faster than $\phi^4$ theory, there is a huge possibility that classicalization might not be able to self-renormalize this type of non-renormalizable (in the Wilsonian sense) theory. Further inquiry in this direction has been left for future work.

\section{Concluding Remarks}

The traditional approach to deal with the theories which seemingly violate unitarity at trans-cut-off energy scattering is to UV-complete them by integrating in new degrees of freedom. A new approach that has been suggested in \cite{dvali2011uv, dvali2011dynamics}, where the theory self-completes itself by forming classical structures at high energy scattering events is called classicalization and is the topic of this paper. The comparative analysis of scattering of monochromatic waves in both theories indicate that $\phi^4$ theory shows typical behaviour of traditional UV-complete theories: the more we increase the energy the smaller length we are able to probe. However, the derivatively coupled theory shows the opposite behaviour which is well suited for classicalization: the scattering event actually is bounded by a characteristic length scale proportional to energy. The subsequent analysis of a collapsing wave packet in the derivatively coupled scalar theory also strengthens the possibility of this classicalization radius ($r_*$) phenomenon. However, although different indicators from different phenomena have been suggested for classicalization, till to date no specific dynamic mechanism has been suggested about how a two particle scattering will produce a large number of particles necessary for classicalization and the feasibility of this $2\rightarrow N$ scattering. In this paper we have focused on this issue. We have calculated the $2\rightarrow N$ scattering amplitude, scattering cross-section and amplitude of classical structure formation for both $\phi^4$ and $\frac{1}{M_*^4}((\partial_\mu\phi)^2)^2$ theories. As it has been showed before that $\phi^4$ theory doesn't classicalize but $\frac{1}{M_*^4}((\partial_\mu\phi)^2)^2$ theory might, so if we consider the ratios of relevant quantities for $\frac{1}{M_*^4}((\partial_\mu\phi)^2)^2$ and $\phi^4$ theories then this ratio should be greater than 1, if the derivatively coupled one really classicalizes. But we got exactly the opposite result: the ratios are less than 1, and decreases even more with the increase in CM energy. In fact, we can make the ratio arbitrarily small by increasing the energy more and more, if we use Dvali's estimate for the number of particles $N$. [The ratio should be more and more greater than 1 if classicalization really happens, because the higher scattering energy is, the more severe the unitarirty violation becomes, the necessity of classicalization increases even more.] So we got negative result from our analysis: the derivatively coupled scalar theory has lesser probability of $2\rightarrow N$ scattering and classical structure formation and the more we increase the CM energy the difference increases even more. Other sources of criticism \cite{akhoury2011no, kovner2012classicalization} regarding classicalization also exist. It might be concluded that, despite having some positive indications, it is very much possible that classicalization might not serve as the UV-completion for this class of non-renormalizable theories. A detail and concrete solution to this open problem is left for future work.

\bibliographystyle{unsrt}  
\bibliography{references}  

\begin{thebibliography}{1}

\bibitem{dvali2011uv}
Gia Dvali, Gian~F Giudice, Cesar Gomez, and Alex Kehagias.
\newblock U{V}-completion by classicalization.
\newblock {\em Journal of High Energy Physics}, 2011(8):108, 2011.

\bibitem{dvali2011dynamics}
Gia Dvali and David Pirtskhalava.
\newblock Dynamics of unitarization by classicalization.
\newblock {\em Physics Letters B}, 699(1-2):78--86, 2011.

\bibitem{dvali2017strong}
Gia Dvali.
\newblock Strong coupling and classicalization.
\newblock In {\em THE FUTURE OF OUR PHYSICS INCLUDING NEW FRONTIERS:
  Proceedings of the International School of Subnuclear Physics}, pages
  189--200. World Scientific, 2017.

\bibitem{goldberg1990breakdown}
Haim Goldberg.
\newblock Breakdown of perturbation theory at tree level in theories with
  scalars.
\newblock {\em Physics Letters B}, 246(3-4):445--450, 1990.

\bibitem{cornwall1993functional}
John~M Cornwall and George Tiktopoulos.
\newblock Functional {S}chr{\"o}dinger equation approach to high-energy
  multileg amplitudes.
\newblock {\em Physical Review D}, 47(4):1629, 1993.

\bibitem{akhoury2011no}
Ratindranath Akhoury, Shinji Mukohyama, and Ryo Saotome.
\newblock No {C}lassicalization {B}eyond {S}pherical {S}ymmetry.
\newblock {\em arXiv preprint arXiv:1109.3820}, 2011.

\bibitem{kovner2012classicalization}
Alex Kovner and Michael Lublinsky.
\newblock Classicalization and unitarity.
\newblock {\em Journal of High Energy Physics}, 2012(11):30, 2012.

\end{thebibliography}






\end{document}